\documentclass[fleqn,10pt]{wlscirep}
\usepackage[utf8]{inputenc}
\usepackage[T1]{fontenc}
\usepackage{svg}
\usepackage{graphicx}
\usepackage{subcaption}
\title{One Giant Leap for Womankind: First Menstrual Cups Tested in Space Flight Conditions}

\author[1,2,3,*,+]{Lígia F. Coelho}
\author[4,+]{Catarina Miranda}
\author[5,+]{João Canas}
\author[4,+]{Miguel Morgado}
\author[4,+]{Diogo Nunes}
\author[6]{André F. Henriques}
\author[1,2,3,7]{Adam B. Langeveld}
\affil[1]{Carl Sagan Institute, Cornell University, Ithaca, NY 14850, USA}
\affil[2]{Department of Astronomy, Cornell University, Ithaca, NY 14850, USA}
\affil[3]{Cornell Center for Astrophysics and Planetary Science, Ithaca, NY 14850 USA}
\affil[4]{Instituto Superior Técnico, Universidade de Lisboa, Lisboa, 1049-001, Portugal}
\affil[5]{Spin.Works S.A., Lisboa, 1700-239, Portugal}
\affil[6]{cE3c—Center for Ecology, Evolution and Environmental Changes, CHANGE -Global Change and Sustainability Institute, Faculdade de Ciências, Universidade de Lisboa, Lisboa, 1749-016 Portugal}
\affil[7]{Department of Physics and Astronomy, Johns Hopkins University, Baltimore, MD 21218, USA}

\affil[*]{Corresponding author: lc992@cornell.edu}

\affil[+]{These authors contributed equally to this work}

\begin{abstract}
In the early days of space exploration, when Sally Ride was offered ``100 tampons'' for a week-long mission, menstrual medical devices first began to be used in space conditions. Since then, hormonal menstrual suppression has become the preferred method for managing menstruation in space, offering significant advantages. However, this is not an option for astronauts who choose to menstruate. The lack of sustainable menstrual technologies will pose challenges for long-duration missions to the Moon and Mars, where astronauts may spend years in space. The AstroCup mission represents the first effort to test menstrual cups in spaceflight, evaluating their durability and functionality. Through material integrity tests and functional assessments using a rheological analogue of human blood, we demonstrate the resilience of menstrual cups and discuss their implications for sustainable menstrual management in future lunar and Martian missions.

\end{abstract}
\begin{document}

\flushbottom
\maketitle
\thispagestyle{empty}

\section*{Introduction}

\begin{figure}[h!]
    \centering
    \includegraphics[width=1\linewidth]{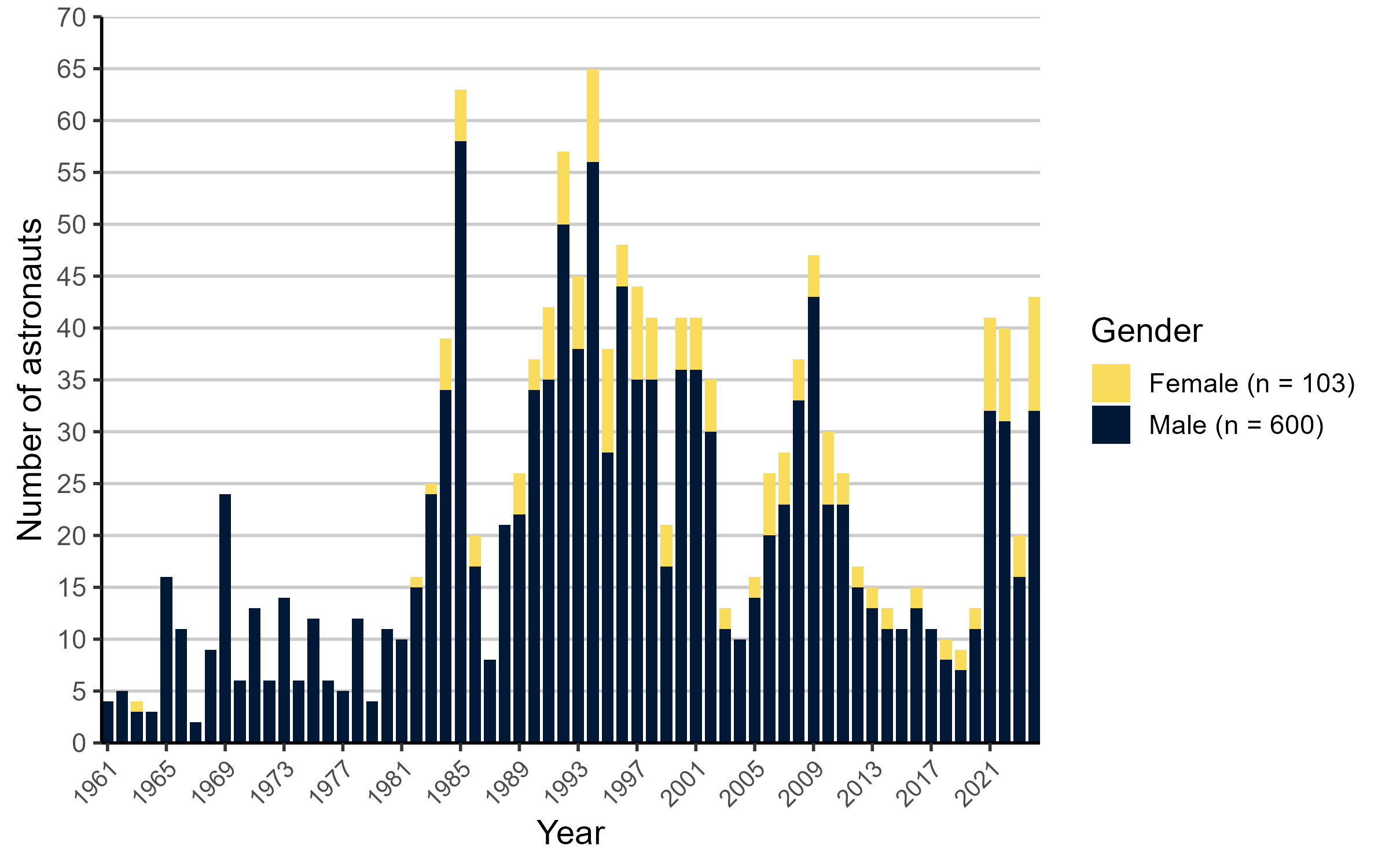}
    \caption{Astronaut launched to space by year per gender since 1961. The first woman flew to space in 1963. Since then, 102 more women have flown to space, representing more than 15\% of total astronauts, in more than 173 launches. The number of astronauts includes reflights by the same person, but in different years. The definition of astronaut is someone who was launched or has flown with the intention of reaching an altitude greater than the Kárman line (including, thus, the ill-fated astronauts from the Challenger disaster, but not the space travellers of Virgin Galactic flights, for example). Data until May 2025 is available in Table S1. 
    Data from:
    \href{https://aerospace.csis.org/}{https://aerospace.csis.org/}, 
    \href{https://www.spacefacts.de/}{https://www.spacefacts.de/}, 
    \href{https://www.nasa.gov/}{https://www.nasa.gov/} 
    and \href{https://en.wikipedia.org}{https://en.wikipedia.org}.
    }
    \vspace{1em}
    \label{fig:fig1}
\end{figure}

In 1963, the first woman ventured into space. Since then, many others have followed suit, with some embarking on extended stays. So far, 103 female astronauts have flown to space in a total of more than 173 launches, including reflights (\autoref{fig:fig1}), more than double the number of women in space before 2000 \cite{jennings_gynecological_2000}.

Astronauts can menstruate in space and microgravity conditions. The first reports of menstrual devices in space begin with the historical first flight of Sally Ride on the Shuttle in 1983. In the next flights, there are reports of astronauts having access to menstrual pads and tampons, and for EVA and launch/landing, all astronauts can use a maximum absorbency garment (MAG) that can retain up to 2000 mL of urine, blood, or feces \cite{jennings_gynecological_2000}.

Currently, hormonal contraception to suppress menstruation is the preferred method to manage menstruation in space\cite{jain2016medically}. Use of hormonal contraceptives in space has several advantages, including practical considerations in waste disposal \cite{steller_menstrual_2021} and the challenges of personal hygiene and pregnancy, since it is contraindicated for pre-flight training activities, including exposure to vacuum chambers or high-performance jet flying \cite{risin2013biomedical, jennings_gynecological_2000}. Hormonal suppression can also help astronauts manage symptoms of some gynecological conditions as well as abdominal pain and bleeding irregularities \cite{risin2013biomedical, jennings_gynecological_2000}. 
Historically, there are parallels in remote and male-dominated environments, such as Antarctic expeditions or military deployments, where studies have shown that the use of menstruation suppression is also common practice \cite{nash2023breaking, jain2016medically}. 

Recent studies have raised questions about the pharmacokinetics of hormonal medications in the space environment and the recent documentation on the increased risk of thromboembolism and reduced bone density in crew members due to combined hormonal suppression drugs \cite{mathyk_understanding_2024, gimunova_effect_2024, steller_gynecologic_2020, jain2016medically}. In fact, the new NASA Decadal Survey (2023) highlights key research areas for the next decade to improve our knowledge of female biology and, more specifically, of the risk of menstrual suppression through oral contraceptives\cite{committee_on_biological_and_physical_sciences_research_in_space_20232032_thriving_2023}. Adding the long-duration mission factor, the consequences of chronic hormonal contraception to astronauts' health are not well studied. Logistically, hormonal suppression drugs can also be an energy burden in space due to the requirement of refrigeration, especially in longer missions, where energy will be increasingly more valuable. Lastly, hormone suppression does not satisfy the needs of astronauts who wish to menstruate. 

\begin{figure}[h!]
    \centering
    \includegraphics[width=1\linewidth]{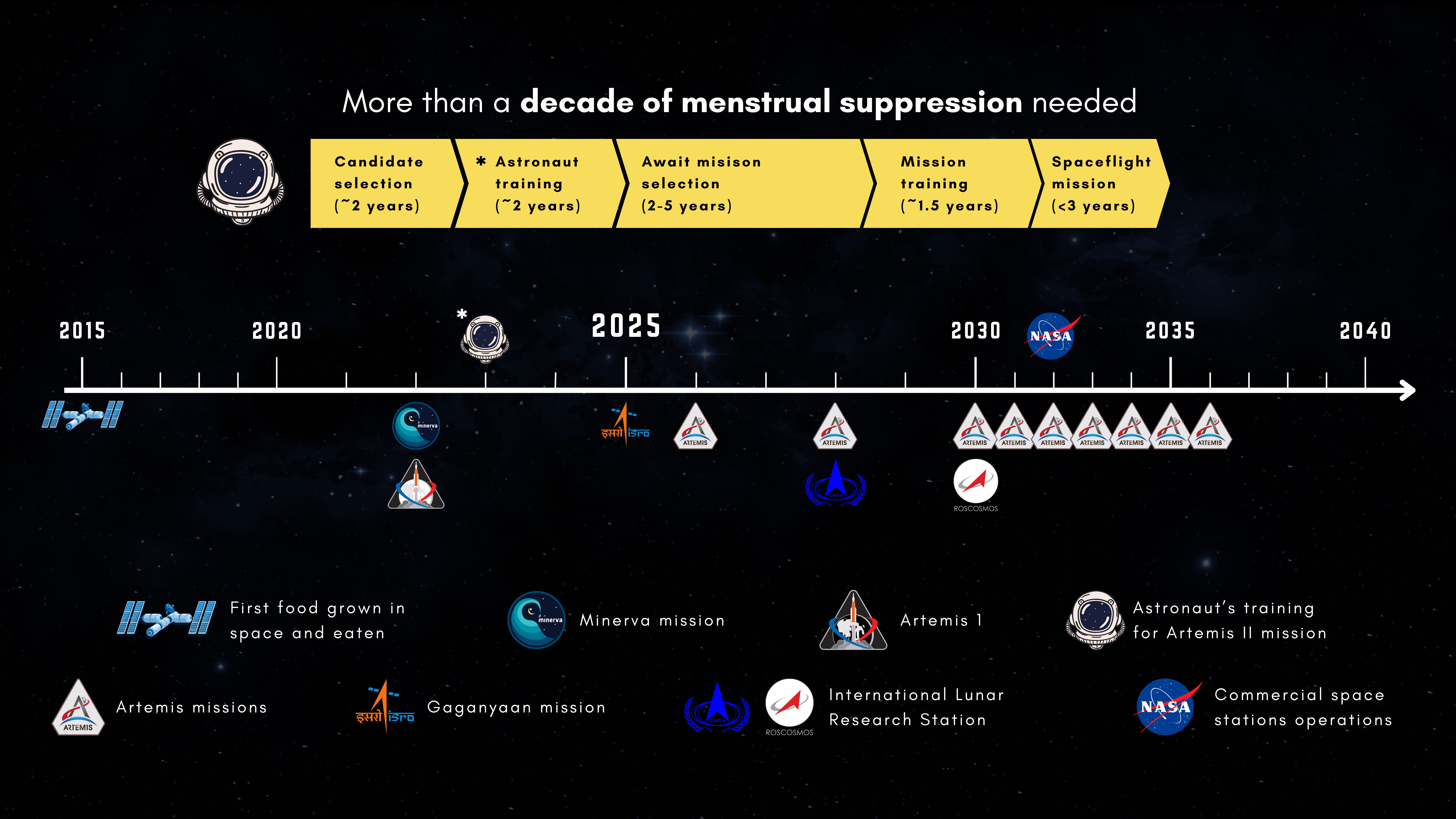}
    \caption{Chronogram of Artemis missions through 2040, highlighting the implications for female astronauts regarding menstrual cycle suppression. An astronaut joining the 2026 Artemis mission and participating in two subsequent Moon missions by 2028 would require over a decade of menstrual suppression. For longer commitments, such as an astronaut involved in the Artemis program from 2025 to 2035, this could extend to nearly 20 years.}
    \label{fig:fig2}
\end{figure}

Currently, 6-month missions are common and still considered long-duration flights. But in the near future, we are expecting missions to be longer than 1 year, possibly going up to decades (\autoref{fig:fig2}) in space when considering the Moon and Mars missions\cite{smith2020artemis}, and this could mean decades of hormone suppression \cite{jain2016medically}. More women will have the opportunity to go to space for even longer missions, and it is paramount that their autonomy on menstrual options is respected. Astronauts on Moon and Mars missions may decide that they would like to keep menstruating for personal preference, as well as for health or reproductive reasons. Female astronauts delay parenthood, on average, by 5.6 years compared to males \cite{gimunova_effect_2024}. For longer missions with the current lack of sustainable menstruation options, this gap may increase drastically. It is crucial to create sustainable options for menstruation in space, especially in preparation for longer missions (\autoref{fig:fig2}).

Studies are not conclusive on the long-term effects of microgravity on the menstrual cycle. The data from astronauts who chose to menstruate in space or who did not take hormonal contraceptives on previous flights is limited \cite{jennings_gynecological_2000, schirmer_1992, mathyk_understanding_2024, risin2013biomedical, gimunova_effect_2024, steller2021menstrual}. Comfortable and sustainable methods for menstruation in space would increase human data on hormonal changes, cyclic menstrual changes, and the complex menstrual cycle dynamics in the space ``exposome'', which is defined as ``the cumulative measure of environmental influences and associated biologic responses throughout the life span, including exogenous exposures and endogenous processes''\cite{miller_nature_2014}. 

Menstrual devices that can be reused or incorporated into normal astronaut clothing will give astronauts options that can solve the logistical burden of common menstrual medical devices, such as menstrual pads or tampons. Currently, there is a lack of available data to drive advances in the development of such medical devices specifically designed for women for use in space missions. Menstrual cups could emerge as a viable solution due to their extensive use on Earth and reusability potential. Similarly to other menstruation-related materials, menstrual cups have not been tested in space.

Research payloads have been one of the main drivers of space medicine, with multiple experiments sent every year to test the limits of medical devices and therapeutics. These include suborbital and orbital flights, the latter including long-term research at the ISS. Several space-related conditions can influence the state and durability of medical devices and therapeutics in space, thus tests for the effect of space launch, turbulence, microgravity, and radiation should be conducted on all proposed materials in preparation for long-term missions where humans need to build as much self-sufficiency as possible. 

Here we present the data of AstroCup, a pioneering research payload to test the compatibility of menstrual cups in space launches.

\section*{Results \& Discussion}

The AstroCup PocketSat mission aimed to test the performance of menstrual cups both before and after the flight, and to analyze the environmental variations the cups experienced during the flight and the effect of the flight in their material integrity and functionality. 

\subsection*{PocketSat mission and metadata}

\begin{figure}[h!]
    \centering
    \vspace{-3em}
    \includegraphics[width=0.85\linewidth]{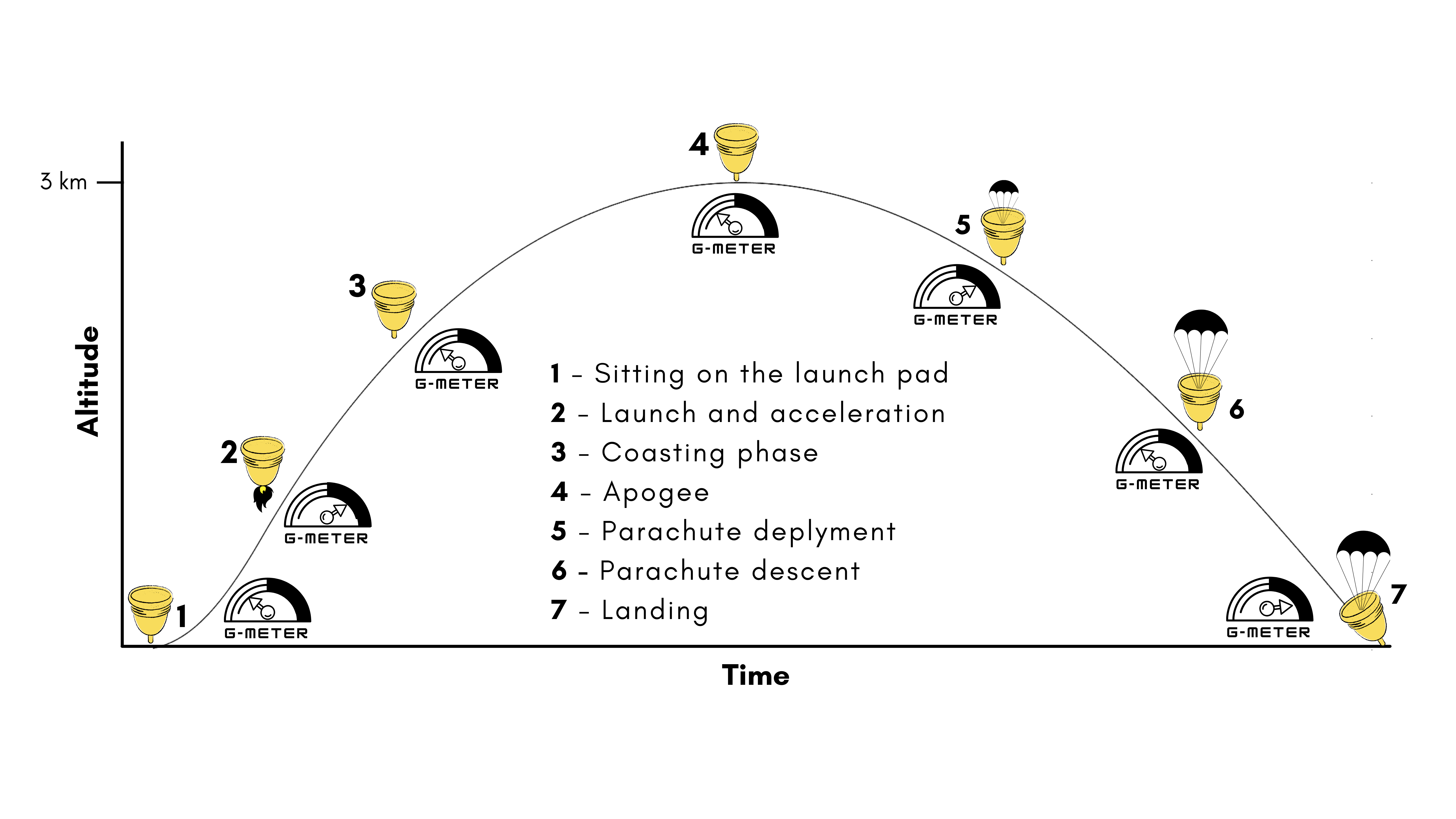}
    \vspace{-3em}
    \caption{Diagram of the flight of the Baltasar rocket with the menstrual cups payload from launch to landing. The key stages of flight are shown, with a "G-meter" highlighting the absolute acceleration experienced by the rocket. At phases 2 (motor burn), 5 (parachute deployment), and 7 (landing), forces are the highest. Phase 1 corresponds to the moment of launch. Phase 3 is the lowest acceleration portion of the ascent, corresponding to a coast phase. Apogee, at 3 km of altitude, corresponds to the instant numbered with a 4. Mirroring phase 3, phase 6 is also a coast phase, where the payload is experiencing low g-forces.}
    \label{fig:fig3}
\end{figure}

The payload containing the experimental cups was launched during the European Rocketry Challenge (EuRoC) competition on October 15th, 2022, on the rocket Baltasar, developed and built by the Rocket Experiment Division (RED), from Instituto Superior Técnico, Lisbon. The flight profile of the Baltasar rocket was successfully executed and monitored across seven distinct phases, from launch to landing, as illustrated in \autoref{fig:fig3}. 

The whole flight lasted 558.93 seconds (approximately 9.3 minutes) after launch. Phase 1 marked the moment of launch, initiating the rocket’s ascent with a rapid increase in acceleration. During Phase 2, the motor burn phase, the rocket experienced one of the highest acceleration peaks, as the propulsion system delivered maximum thrust. This was followed by Phase 3, a coast phase characterized by the lowest acceleration during the ascent, where the rocket relied on its momentum with minimal additional thrust. The rocket reached apogee (Phase 4) at 25.75 seconds, achieving an altitude of over 3 km, where the transition from ascent to descent occurred due to the effect of gravity. During the descent, parachute deployment occurred at 30.13 seconds (Phase 5), which induced another peak in acceleration due to the sudden deceleration caused by the parachute’s drag. Phase 6 mirrored Phase 3 as a coast phase, with the rocket descending under parachute and experiencing low g-forces, ensuring minimal stress on the payload. Finally, Phase 7, the landing, recorded a third peak in acceleration as the rocket made contact with the ground, though the parachute mitigated the impact to within acceptable limits for the payload. The data highlighted that the highest forces were experienced during Phases 2, 5, and 7, corresponding to motor burn, parachute deployment, and landing, respectively. These phases were the most likely to subject the menstrual cup payload to significant mechanical stress. The Baltasar rocket maintained structural integrity and payload safety throughout the flight.

\begin{figure}[h!]
    \centering
    \includegraphics[width=1\linewidth]{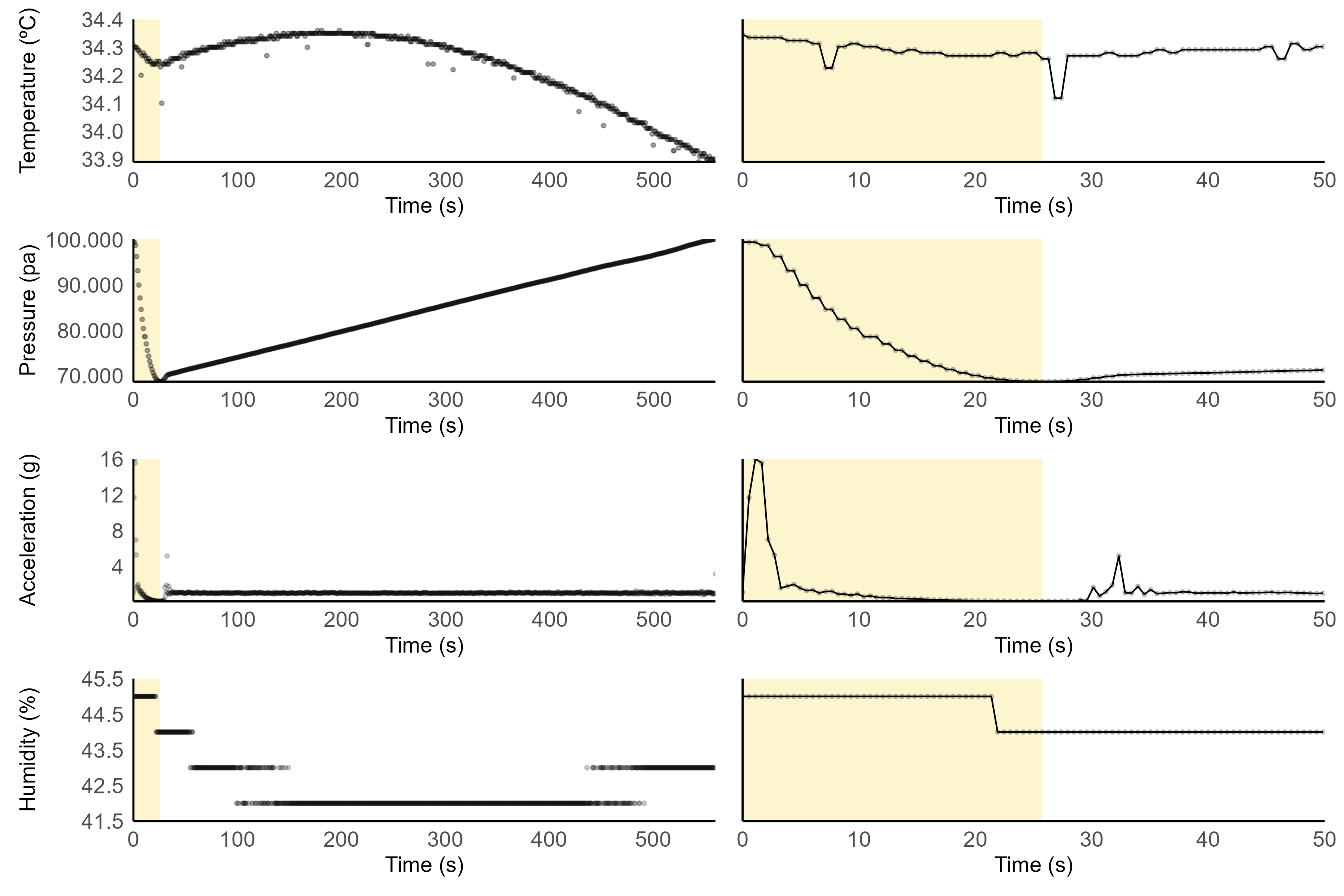}
    \caption{Time-dependent temperature, pressure, acceleration, and humidity conditions experienced by the menstrual cups during flight. The plots on the left show the full flight, starting at lift-off (launch at 0 seconds) and ending when the rocket lands back on the ground. Hence, for clarity, the area of the plot highlighted in yellow is the ascent portion of the flight. The region with no color shows the descent. The plots on the right show the ``zoom'' version focusing only on the first 50 seconds. We can observe that apogee is achieved at 25.75 seconds, and parachute deployment occurs at 31.13 seconds into flight, as evidenced by a local acceleration peak. Landing occurs at 559.93 seconds (approximately 9.3 minutes) after launch.}
    \label{fig:fig4}
\end{figure}

From the moment the rocket systems were powered on until the vehicle was disarmed by the recovery team, the payload computer on board recorded over two and a half hours of data at a frequency of approximately 2 Hz. The electronics monitored several environmental conditions within the payload, crucial for assessing the performance of the menstrual cups during the flight. These conditions included air pressure, temperature, humidity, and linear accelerations (\autoref{fig:fig4}).

During the flight, the barometer registered a pressure drop below 70,000\,($\pm 100.0$)\,Pa, which, according to standard atmospheric model pressures\cite{IntStandardAtmos}, corresponds to an apogee height of approximately 3,024 meters. The recorded temperature inside the payload ranged between 32.5--34.3\,($\pm 1.0)^\circ$C, while humidity gradually decreased to approximately 40\,($\pm 5$)\,\%. The elevated initial temperature within the payload is potentially attributable to exposure to direct sunlight at the launch site before launch (approximately 22$^\circ$C). At an altitude of 3\,km the temperature was close to 6$^\circ$C. 
A critical aspect of the mission was the measurement of the payload's acceleration using the on-board 3-axis Inertial Measurement Unit (IMU). This sensor has a range from -156.8 to +156.8\,m/s$^2$. During the flight, the maximum acceleration exceeded the sensor’s upper limit, which is equivalent to 16 times the Earth's gravitational acceleration. Since the sensor’s upper limit already surpasses the accelerations experienced by most orbital launch vehicles \cite{deepika_spacecraft_2021}, the data (\autoref{fig:fig4}) confirms that the flight provided an appropriate simulation environment for testing the menstrual cups under conditions similar to those of an orbital launch. 

 
\subsection*{Menstrual cups performance}


On October 9th, 2022, 8-hour pre-flight tests assessed the integrity and performance of menstrual cups. Glycerol was used to evaluate performance, while water tested structural integrity. Both liquids performed as expected, with the cups showing no leaks, liquid contamination, or material degradation, confirming their functionality.

After the flight, post-flight performance tests performed on October 15th, 2022, began with a thorough visual inspection of the two flown test cups. This inspection revealed no signs of wear or tear. On October 17th, the post-flight leak tests were conducted. The material integrity and the performance of the flown menstrual cups were consistent with the pre-flight tests. No leaks of glycerol or water were detected in any of the tested menstrual cups. No observable differences compared to the two control units that remained on the ground on the day of the flight (\autoref{fig:fig6}) were reported. Therefore, after enduring the flight conditions on board the Baltasar rocket, the menstrual cups exhibited no performance degradation. The materials and seals remained intact, confirming their resilience and functionality under the stress of spaceflight conditions.

\subsection*{Implications for missions to the Moon and Mars}

These tests confirmed that the menstrual cups maintained their structural integrity and functionality, with no leaks of glycerol or water and no signs of wear or tear. This resilience is critical for space missions, where equipment must endure rigorous conditions. Menstrual cups could be a reliable and sustainable option for managing menstruation during missions to the Moon and Mars. Unlike disposable products, they reduce waste, which is a key advantage in the confined, resource-limited environment of space. This could improve astronauts’ quality of life, autonomy, and simplify logistics for long-term missions.

The experiment was conducted within Earth’s atmosphere at 3 km altitude, while the Moon and Mars present distinct gravitational and atmospheric conditions. The Moon has about 1/6th of Earth’s gravity and no atmosphere. The reduced gravity on the Moon might affect how menstrual fluid flows upon the removal of the cup. The same rationale applies to the ISS. Mars has about 1/3rd of Earth’s gravity and a thin atmosphere. These conditions are closer to Earth’s, but differences in gravity and pressure could still influence performance. The success of AstroCup in Earth’s gravity is a positive starting point, but further testing is needed to confirm how the cups perform in reduced gravity. For example, lower gravity might alter fluid dynamics, potentially requiring adjustments to the cups’ design or usage. Simulations in microgravity (e.g., via parabolic flights) would help verify their effectiveness for the Moon and Mars.
This flight was a short-duration test, while missions to the Moon (days) or Mars (months) are much longer. The cups’ performance over a single flight is encouraging, but extended use that include multiple menstrual cycles, cleaning, and storage, remains untested in the experiment. For lunar and Martian missions, long-duration studies are essential to ensure the cups remain functional and hygienic over time. This would involve testing their durability and performance across weeks or months in space-like conditions or analogue human missions on Earth. Space exposes equipment to radiation, extreme temperatures, and vacuum conditions. These factors were not tested in this setup, but they could affect the menstrual cups’ materials (e.g., silicone) or seals over time. 
Other menstrual devices like menstrual underwear should also be considered and tested. Currently, underwear in space is treated as a consumable due to the lack of technology to clean clothes. Menstrual underwear could be a readily available solution for menstruation in current and future missions, but requires testing to know if it can be repurposed or needs adaptations. Menstrual devices need further testing in space conditions if women are to have a choice to menstruate during long missions to the Moon and Mars.

\section*{Conclusions and Future perspectives}

Planning long missions to the Moon and Mars requires advancements in technologies relating to human-flight and life-support systems, including medical devices tested and adjusted to space conditions. There is a lack of research and tests of sustainable and reusable menstrual devices under space conditions. This lack of research hinders the availability, adjustment, or incorporation of solutions and options to menstruate in space. Menstruation in space needs to be accessible and easy for everyone. This is a crucial issue for the autonomy and health of astronauts and will become more apparent as longer missions take place. 
Menstrual devices should be incorporated within future payloads and missions that include tests for the endurance of the material to microgravity and radiation exposure over time, the efficacy of the devices in the space environment over time used by menstruating astronauts, and the sanitation of the use of devices and disposable techniques. 
The AstroCup project is the first step in the study of the endurance and feasibility of using menstrual devices in space, and this experiment proves that these cups withstand the launch environment. Lastly, this experiment shows that, more than launching menstrual cups, it is also important to launch the discussion about choice and autonomy in healthcare in space.

\section*{Methods}

\begin{figure}[h!]
    \centering
    \includegraphics[width=0.85\linewidth]{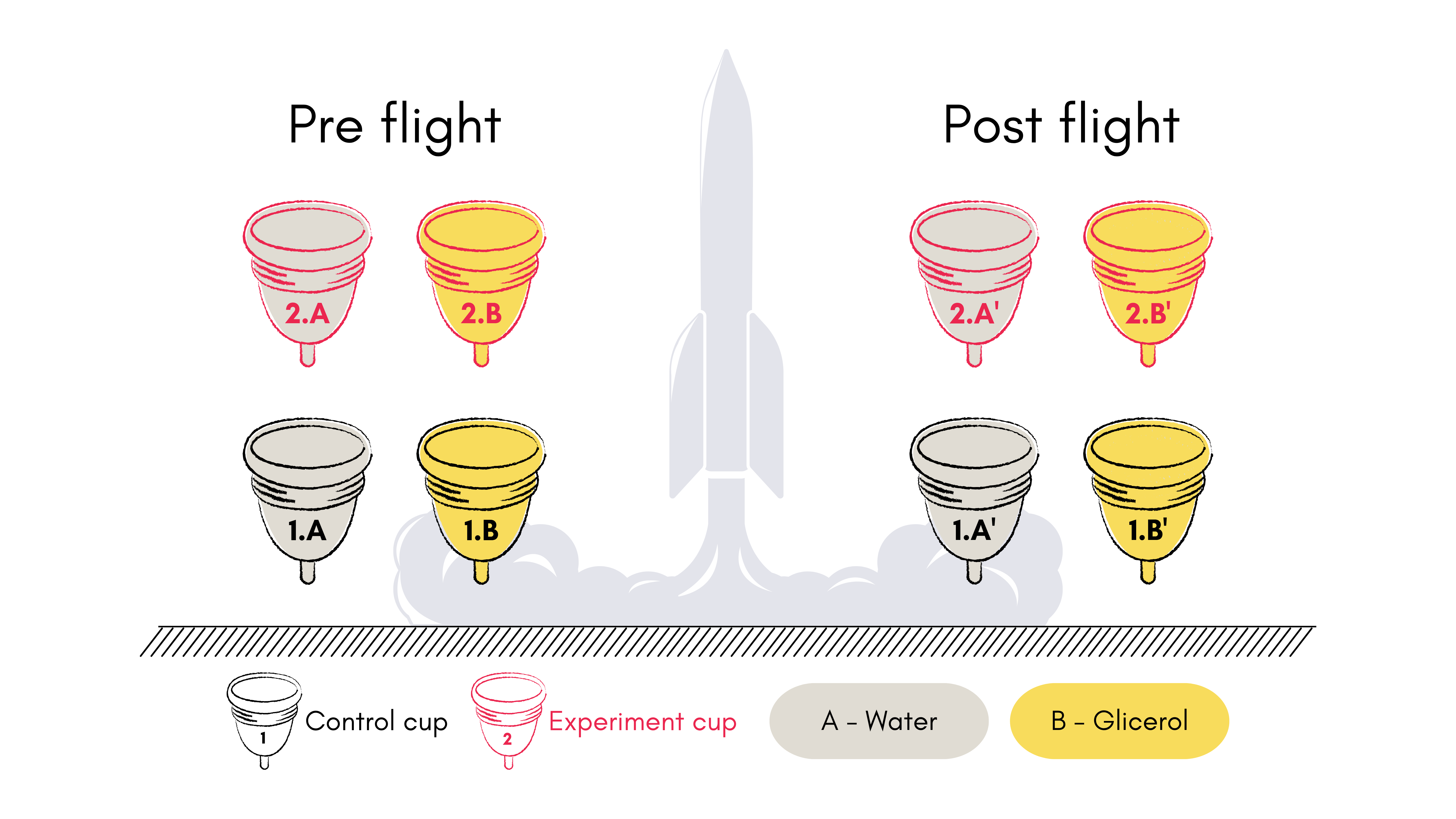}
    \caption{Diagram of experiments conducted on menstrual cups before launch (left) and after flight (right). Ground control cups are outlined in black, while experimental cups are outlined in pink. Gray and yellow filled colors indicate tests using water (to assess material integrity) and glycerol (a blood analogue), respectively.}
    \label{fig:fig6}
\end{figure}

\begin{figure}[h!]
    \centering
    \vspace{-3.5em}
    \includegraphics[width=1\linewidth]{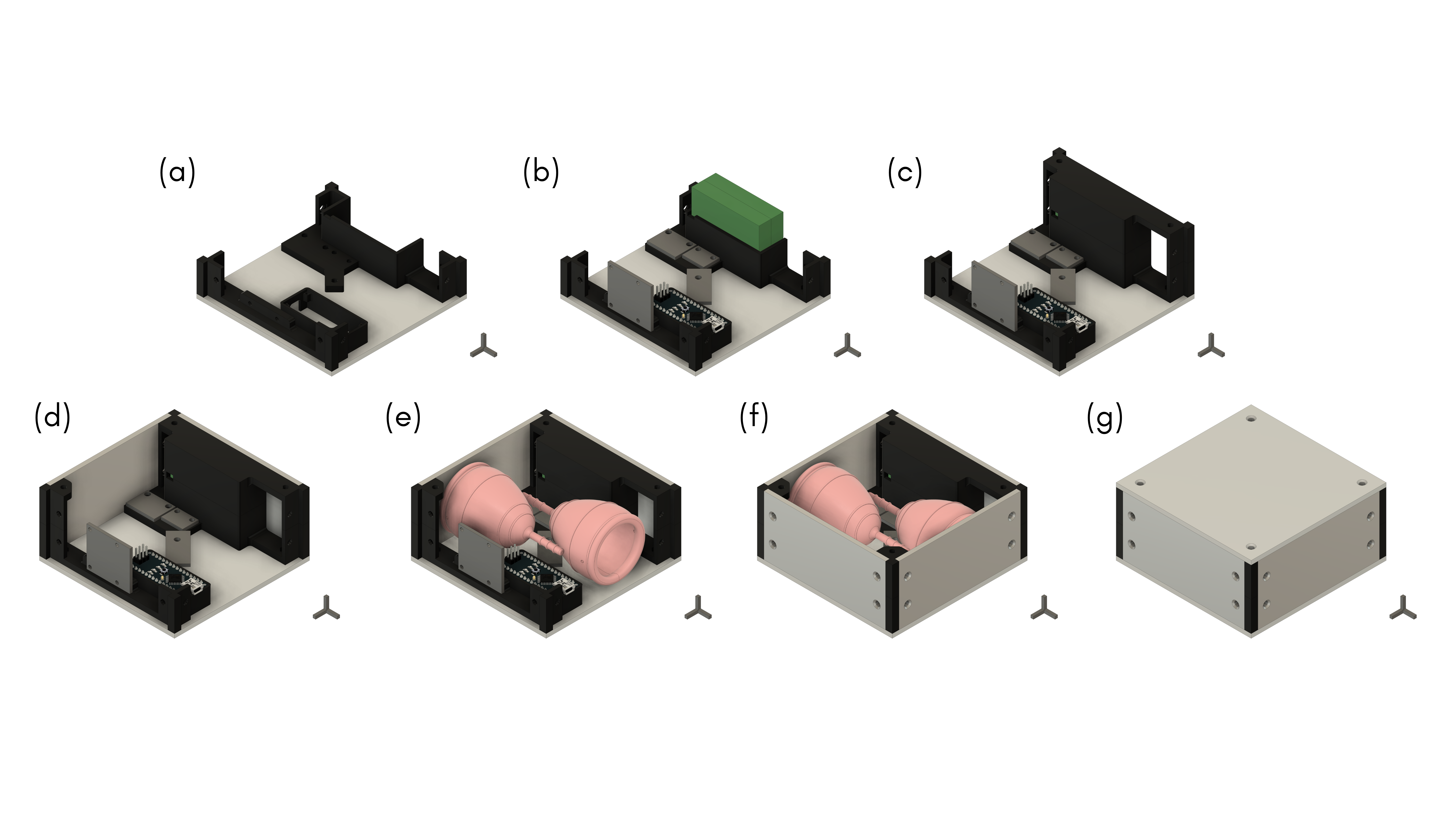}
    \vspace{-3em}
    \caption{Scheme of payload assembly process, from step (a) to (g). (a) the first piece is called the baseframe and it already contains most of the internal plastic structures; (b) incorporation of batteries and electronics; (c) placement of battery enclosure; (d) closure of two sides; (e) placement of the menstrual cups inside the payload box; (f) closure of two other sides; (g) the side panels and top lid are closed using 4 M3 countersunk hex screws on the four corners.}
    \vspace{0.5em}
    \label{fig:fig5}
\end{figure}

The payload was prepared for launch on October 14th, 2022, the day before EuRoC. This preparation involved sealing the payload, which included batteries and the menstrual cups for testing, and integrating it into RED's payload adapter.
The experimental design included four commercial menstrual cups (manufactured by Lunette). Two cups served as ground controls, and the other two were sent in a payload on the rocket (\autoref{fig:fig5}). All cups underwent a series of tests prior (pre-flight performance tests) and after (post-flight performance tests) the flight to assess any changes in their functionality and integrity (\autoref{fig:fig6}).

The ground control cups were kept under standard laboratory conditions to provide a baseline reference for comparison. These controls allowed researchers to isolate the effects of the flight environment on the flown cups. The two experimental cups were integrated into the AstroCup payload, which was then installed on the Baltasar rocket.

The payload containing the experimental cups was launched during the EuRoC competition on October 15th, 2022. Throughout the flight, various environmental parameters, including air pressure, temperature, humidity, and linear accelerations, were recorded to understand the conditions the menstrual cups endured (\autoref{fig:fig4}). All statistical analyses were conducted in R (v.4.5.0) using a custom
script, with the following packages: tidyverse
(v.2.0.0), scales (v.1.4.0), patchwork (v.1.3.0), ggplot2 (v.3.5.2),
and dplyr (v.1.1.4).
Upon the rocket's return, the menstrual cups were inspected on site for any visible deformations and after passing that initial test, post-flight tests were conducted on both the experimental and control cups to identify any potential performance degradation or material changes. These results of the experimental cups were then compared to the pre-flight tests and the ground control results to measure the menstrual cups' resilience and performance under the extreme conditions of a rocket flight.

\subsection*{The Experiment}

In each pair of menstrual cups (test vs control), one of the cups was tested with water and the other with a glycerol solution (\autoref{fig:fig6}). Water is a neutral liquid, less viscous than glycerol allowing analysis of any damages to the structure of the cup. A glycerol solution (50\,\% diluted in water) is a standard rheology analog to human blood \cite{thurston1990rheological,risin2013biomedical}. In both tests with liquids (pre-flight and post-flight -- \autoref{fig:fig6}), 25\,mL of water and then 25\,mL of glycerol solution was decanted in each cup and left to sit for 8-12 hours, which corresponds to a standard period of menstrual cup usage considering data from the Screening Life Cycle Assessments (LCA): Reusable Menstrual Cup in Europe by the European Comission in 2022.

\subsection*{Payload Design}

A payload casing was designed and built to transport menstrual cups, as well as house the electronic systems needed to monitor flight conditions. Most of the design decisions were driven by the size and weight constraints of the sounding rocket's payload bay, which limited the cargo to a standard 4U PocketSat. A 4U PocketSat has a size of 100x100x50 mm. The menstrual cup (Lunette) measured 41 mm in diameter and 72 mm in length, which means two cups can fit inside the volume of the PocketSat with enough leftover volume for the electronic and structural components. 

Additionally, it was required for the payload to weigh between 800 and 1000 g. Considering the low density and weight of the cups (14 g each), the employment of ballast was deemed necessary. However, instead of placing ballast, the team decided to add additional weight to the payload structure by constructing it using steel instead of 3D printed plastic. Since there are no wireless communications on the system, a thick metal structure can be used to add mass to the system. The structure CAD models can be observed in Figure (\autoref{fig:fig5}). The payload walls are made of steel sheets 3 mm thick. The corners of the structure's body are 3D-printed in PLA. The payload lid is closed on the four corners using 4 M3 countersunk hex screws.
 
\section*{Acknowledgements}
LFC is funded by the Heising-Simons Foundation 51 Pegasi b postdoctoral fellowship 2024-5174. The authors gratefully acknowledge the Rocket Experiment Division (RED) and the Aerospace Engineering Students' Organization (AeroTéc) from Instituto Superior Técnico, for all work and data related to the Baltasar rocket. We extend our sincere appreciation to the RED team for selecting our proposal, which made the menstrual cups flight experiment possible. Catarina Pereira is gratefully acknowledged for her significant contributions to the study's conceptualization, logistics, and experimental work. The authors acknowledge the European Rocket Challenge (EuRoC). The authors acknowledge Lunette for donating the menstrual cups for the study.

\bibliography{sample}

\begin{thebibliography}{10}
\urlstyle{rm}
\expandafter\ifx\csname url\endcsname\relax
  \def\url#1{\texttt{#1}}\fi
\expandafter\ifx\csname urlprefix\endcsname\relax\def\urlprefix{URL }\fi
\expandafter\ifx\csname doiprefix\endcsname\relax\def\doiprefix{DOI: }\fi
\providecommand{\bibinfo}[2]{#2}
\providecommand{\eprint}[2][]{\url{#2}}

\bibitem{jennings_gynecological_2000}
\bibinfo{author}{Jennings, R.~T.} \& \bibinfo{author}{Baker, E.~S.}
\newblock \bibinfo{journal}{\bibinfo{title}{Gynecological and {Reproductive} {Issues} for {Women} in {Space}: {A} {Review}}}.
\newblock {\emph{\JournalTitle{Obstetrical \& Gynecological Survey}}} \textbf{\bibinfo{volume}{55}} (\bibinfo{year}{2000}).

\bibitem{jain2016medically}
\bibinfo{author}{Jain, V.} \& \bibinfo{author}{Wotring, V.~E.}
\newblock \bibinfo{journal}{\bibinfo{title}{Medically induced amenorrhea in female astronauts}}.
\newblock {\emph{\JournalTitle{npj Microgravity}}} \textbf{\bibinfo{volume}{2}}, \bibinfo{pages}{1--6} (\bibinfo{year}{2016}).

\bibitem{steller_menstrual_2021}
\bibinfo{author}{Steller, J.~G.} \emph{et~al.}
\newblock \bibinfo{journal}{\bibinfo{title}{Menstrual management considerations in the space environment}}.
\newblock {\emph{\JournalTitle{REACH}}} \textbf{\bibinfo{volume}{23-24}}, \bibinfo{pages}{100044}, \doiprefix\url{10.1016/j.reach.2021.100044} (\bibinfo{year}{2021}).

\bibitem{risin2013biomedical}
\bibinfo{author}{Risin, D.} \& \bibinfo{author}{Stepaniak, P.~C.}
\newblock \emph{\bibinfo{title}{Biomedical results of the space shuttle program}}.
\newblock \bibinfo{number}{NASA/SP-2013-607} (\bibinfo{publisher}{National Aeronautics and Space Administration}, \bibinfo{year}{2013}).

\bibitem{nash2023breaking}
\bibinfo{author}{Nash, M.}
\newblock \bibinfo{journal}{\bibinfo{title}{Breaking the silence around blood: managing menstruation during remote antarctic fieldwork}}.
\newblock {\emph{\JournalTitle{Gender, Place \& Culture}}} \textbf{\bibinfo{volume}{30}}, \bibinfo{pages}{1083--1103} (\bibinfo{year}{2023}).

\bibitem{mathyk_understanding_2024}
\bibinfo{author}{Mathyk, B.} \emph{et~al.}
\newblock \bibinfo{journal}{\bibinfo{title}{Understanding how space travel affects the female reproductive system to the {Moon} and beyond}}.
\newblock {\emph{\JournalTitle{npj Women's Health}}} \textbf{\bibinfo{volume}{2}}, \bibinfo{pages}{20}, \doiprefix\url{10.1038/s44294-024-00009-z} (\bibinfo{year}{2024}).

\bibitem{gimunova_effect_2024}
\bibinfo{author}{Gimunová, M.}, \bibinfo{author}{Paludo, A.~C.}, \bibinfo{author}{Bernaciková, M.} \& \bibinfo{author}{Bienertova-Vasku, J.}
\newblock \bibinfo{journal}{\bibinfo{title}{The effect of space travel on human reproductive health: a systematic review}}.
\newblock {\emph{\JournalTitle{npj Microgravity}}} \textbf{\bibinfo{volume}{10}}, \bibinfo{pages}{10}, \doiprefix\url{10.1038/s41526-024-00351-1} (\bibinfo{year}{2024}).

\bibitem{steller_gynecologic_2020}
\bibinfo{author}{Steller, J.~G.} \emph{et~al.}
\newblock \bibinfo{journal}{\bibinfo{title}{Gynecologic {Risk} {Mitigation} {Considerations} for {Long}-{Duration} {Spaceflight}}}.
\newblock {\emph{\JournalTitle{Aerospace Medicine and Human Performance}}} \textbf{\bibinfo{volume}{91}}, \bibinfo{pages}{543--564}, \doiprefix\url{10.3357/AMHP.5538.2020} (\bibinfo{year}{2020}).

\bibitem{committee_on_biological_and_physical_sciences_research_in_space_20232032_thriving_2023}
\bibinfo{author}{{Committee on Biological and Physical Sciences Research in Space 2023–2032}}, \bibinfo{author}{{Space Studies Board}}, \bibinfo{author}{{Division on Engineering and Physical Sciences}} \& \bibinfo{author}{{National Academies of Sciences, Engineering, and Medicine}}.
\newblock \emph{\bibinfo{title}{Thriving in {Space}: {Ensuring} the {Future} of {Biological} and {Physical} {Sciences} {Research}: {A} {Decadal} {Survey} for 2023-2032}} (\bibinfo{publisher}{National Academies Press}, \bibinfo{address}{Washington, D.C.}, \bibinfo{year}{2023}).

\bibitem{smith2020artemis}
\bibinfo{author}{Smith, M.} \emph{et~al.}
\newblock \bibinfo{title}{The artemis program: An overview of nasa's activities to return humans to the moon}.
\newblock In \emph{\bibinfo{booktitle}{2020 IEEE aerospace conference}}, \bibinfo{pages}{1--10} (\bibinfo{organization}{IEEE}, \bibinfo{year}{2020}).

\bibitem{schirmer_1992}
\bibinfo{author}{Schirmer~JU, W.~W.}
\newblock \bibinfo{journal}{\bibinfo{title}{Menstrual history in altitude chamber trainees}}.
\newblock {\emph{\JournalTitle{Aviat Space Environ Med}}} \textbf{\bibinfo{volume}{63}}, \bibinfo{pages}{2} (\bibinfo{year}{2000}).

\bibitem{steller2021menstrual}
\bibinfo{author}{Steller, J.~G.} \emph{et~al.}
\newblock \bibinfo{journal}{\bibinfo{title}{Menstrual management considerations in the space environment}}.
\newblock {\emph{\JournalTitle{Reach}}} \textbf{\bibinfo{volume}{23}}, \bibinfo{pages}{100044} (\bibinfo{year}{2021}).

\bibitem{miller_nature_2014}
\bibinfo{author}{Miller, G.~W.} \& \bibinfo{author}{Jones, D.~P.}
\newblock \bibinfo{journal}{\bibinfo{title}{The {Nature} of {Nurture}: {Refining} the {Definition} of the {Exposome}}}.
\newblock {\emph{\JournalTitle{Toxicological Sciences}}} \textbf{\bibinfo{volume}{137}}, \bibinfo{pages}{1--2}, \doiprefix\url{10.1093/toxsci/kft251} (\bibinfo{year}{2014}).

\bibitem{IntStandardAtmos}
\bibinfo{author}{ISO}.
\newblock \bibinfo{title}{Standard atmosphere}.
\newblock \bibinfo{type}{Standard}, \bibinfo{institution}{International Organization for Standardization} (\bibinfo{year}{1975}).

\bibitem{deepika_spacecraft_2021}
\bibinfo{author}{Deepika, L.}, \bibinfo{author}{Nataraja, M.}, \bibinfo{author}{Mishra, S.} \& \bibinfo{author}{Kumar, A.}
\newblock \bibinfo{journal}{\bibinfo{title}{Spacecraft launch and re-entry: {Effects} of simulated +{Gx} acceleration on cardiorespiratory parameters}}.
\newblock {\emph{\JournalTitle{Indian Journal of Aerospace Medicine}}} \textbf{\bibinfo{volume}{65}}, \bibinfo{pages}{69--73}, \doiprefix\url{10.25259/IJASM_18_2021} (\bibinfo{year}{2021}).

\bibitem{thurston1990rheological}
\bibinfo{author}{Thurston, G.~B.}
\newblock \bibinfo{title}{Rheological analogs for human blood in large vessels}.
\newblock In \emph{\bibinfo{booktitle}{Biofluid Mechanics: Blood Flow in Large Vessels}}, \bibinfo{pages}{367--374} (\bibinfo{publisher}{Springer}, \bibinfo{year}{1990}).

\end{thebibliography}

%

%

\section*{Author contributions statement}

LFC, CM, JC, MM, and DN conceptualized the research. CM, JC, DN, and MM led the experiments. LFC wrote the manuscript. CM, JC, MM, DN, and ABL contributed to the original manuscript draft. DN, CM, AH, and MM led the data treatment. LFC, MM, CM, JC, and DN led the data interpretation. All authors contributed to the manuscript revision.

%


%
%





\end{document}